\documentclass[pre,twocolumn]{revtex4-2}

\usepackage{graphicx}
\usepackage{dcolumn}
\usepackage{bm}

\usepackage[utf8]{inputenc}
\usepackage[T1]{fontenc}
\usepackage{mathptmx}

\usepackage{tikz}

\usepackage{amsmath}
\usepackage{mathtools}
\usepackage{amssymb}

\usepackage{pst-node}
\usepackage{verbatim}   
\usepackage{color}      
\usepackage{subfigure}  
\usepackage{hyperref}   

\usepackage{braket}

\newcommand{\ba}{\begin{eqnarray}}
\newcommand{\ea}{\end{eqnarray}}
\newcommand{\be}{\begin{equation}}
\newcommand{\ee}{\end{equation}}

\pdfoutput=1

\begin{document}

\title{Forward--backward correspondence between stationary structure and splitting probabilities in active matter}

\author{Derek Frydel}
\affiliation{Department of Chemistry, Universidad Técnica Federico Santa María, Campus San Joaquin, Santiago, Chile}

\date{\today}

\begin{abstract}
Active particles confined by hard walls accumulate at boundaries and may become dynamically adsorbed due to directional persistence.
In this work, we show that the same persistence mechanism also gives rise to a finite wall splitting probability, meaning that a particle 
initialized at a wall can reach the opposite wall before returning to its starting point.
By comparing forward and backward evolution equations in position--velocity phase space, we derive exact relations linking 
stationary distributions and splitting probabilities for run-and-tumble, active Brownian, and active Ornstein--Uhlenbeck particles.
In particular, we show that the stationary density equals the spatial derivative of the splitting probability, while the fraction 
of adsorbed particles is encoded in the splitting probabilities evaluated at the walls.
The correspondence is valid in arbitrary spatial dimension and establishes an exact bridge between stationary and first-passage 
descriptions of confined active matter, revealing stationary and first-passage observables as complementary representations of 
the same persistence-driven dynamics.
\end{abstract}

\pacs{
}

\maketitle


\section{Introduction}

A distinctive feature of active particles under confinement is their tendency to accumulate near boundaries.
Under hard-wall confinement and in the absence of thermal fluctuations, 
such accumulation leads to dynamical adsorption at the walls
\cite{PRE-Schnitzer-1993, PRL-Cates-2008, PRL-FilyMarchetti-2012, pressure-2015b, RMP-Bechinger-2016, PRE-FrydelPodgornik-2023, Farago-2024}.
We refer to this phenomenon as \emph{dynamical adsorption} 
because it is a nonequilibrium behavior resulting from the interplay between directional 
persistence and overdamped dynamics, rather than from attractive interactions with the walls.
The resulting stationary state consists of coexisting adsorbed and free particles, whose stationary distribution is governed 
by the \emph{forward} Kolmogorov (Fokker--Planck) equation.

Beyond stationary structure, another important class of observables describing confined active particles is provided by first-passage quantities
\cite{Redner2001,Oshanin-2012,AP-Bray-2013}.
For active particles, survival probabilities, exit statistics, and first-passage times have been studied extensively in recent years
\cite{Fedotov-PRE-2010,Malakar-2018,PRL-Mori-2020,Basu2019,POF-Frydel-2024,book-Basu-2024}.
Unlike stationary distributions, these quantities are governed by the \emph{backward} Kolmogorov equation
\cite{Redner2001,Demaerel2018}.

The focus of the present work is the splitting probability $\pi_L(x)$, defined as the probability that a particle initially located at 
$x\in(0,L)$ reaches the wall at $x=L$ before reaching the wall at $x=0$.
Our interest in this problem was motivated by the recent work of Klinger \textit{et al.}~\cite{PRL-Klinger-2022}, who showed that a discontinuous random 
walk, namely a jumping-particle process, exhibits the condition $\pi_L(0)>0$.
This result is striking because standard Brownian motion satisfies $\pi_L(0)=0$. In other words, a particle initially located at the wall 
can never reach the opposite wall before returning.

In the present work, we show that discontinuous random-walk dynamics are not essential for the condition $\pi_L(0)>0$.
Continuous random walks with directional persistence, in the absence of thermal fluctuations, also satisfy this condition.
This suggests that discontinuous random walks belong to a broader class of random walks with directional persistence.
Since the same directional persistence also gives rise to dynamical adsorption, it is natural to ask whether dynamical adsorption 
and the condition $\pi_L(0)>0$ are connected, and whether this connection admits a rigorous mathematical formulation.

Since dynamical adsorption and the splitting probability are governed by the forward and backward Kolmogorov equations, respectively, 
establishing a rigorous connection between them requires a direct comparison of these two formulations. 
One of the main objectives of this work is to reformulate the forward and backward problems in forms that admit a direct transformation 
between them, thereby relating their respective solutions.

For a jumping particle confined between two walls, the stationary distribution and the splitting probability are formulated as integral 
recurrence equations rather than differential equations. By comparing the two recurrence formulations, it was possible to establish 
an exact relation between the fraction of adsorbed particles and the wall splitting probability, $\pi_L(0)$~\cite{POF-Frydel-2024}. 
Since run-and-tumble particles (RTPs) can be mapped onto jumping particles by averaging over the swimming velocity, this result 
provided the first indication that a similar correspondence might exist more generally. Independently, Guéneau and Touzo established 
the same correspondence for one-dimensional RTPs using a differential-equation formulation~\cite{JPA-Gueneau-2024}.

After deriving the correspondence presented here, we recognized that it can be viewed as part of the broader framework of stochastic dualities. 
More specifically, it is closely related to the Siegmund duality introduced by David Siegmund in 1976~\cite{AP-Siegmund-1976}, 
which establishes an equivalence between absorbing and reflecting barrier problems. Guéneau and Touzo subsequently generalized this framework 
to a broad class of stochastic processes~\cite{JSTAT-Gueneau-2024}, and later applied it to active Brownian particles in two spatial 
dimensions~\cite{Baouche-2026-archive}.

The purpose of this work is to establish an exact relation between the wall splitting probability and the fraction of adsorbed 
particles for the canonical models of active matter: run-and-tumble particles (RTPs), active Brownian particles (ABPs), 
and active Ornstein–Uhlenbeck particles (AOUPs). Since these dynamics are Markovian only in the extended phase space \((x,v)\), 
where \(v\) denotes the component of the swimming velocity normal to the confining walls,
we formulate the correspondence directly in this space. This yields a unified framework applicable to all three models and, 
more generally, to arbitrary spatial dimensions.

This paper is organized as follows. In Sec.~\ref{sec:trajectories}, we discuss the short-time origin of finite wall splitting probabilities.
In Sec.~\ref{sec:framework}, we formulate the forward and backward problems in a common position--velocity phase space and derive 
the correspondence between stationary structure and splitting probabilities.  In Sec.~\ref{sec:results}, we present the resulting identities, 
specify the model-dependent velocity distributions for RTP, ABP, and AOUP dynamics, and discuss the connection with jumping-particle processes.
In Sec.~\ref{sec:numerics}, we verify the theoretical predictions numerically.  Finally, we conclude with a summary of the main results and their implications.

\section{Short-time origin of finite wall splitting probabilities}
\label{sec:trajectories}

Before developing the forward--backward formulation presented in the next section, we first examine the microscopic origin 
of the condition \(\pi_L(0,v)>0\).  We consider two parallel walls located at $x=0$ and $x=L$.  Although the system may be 
embedded in an arbitrary spatial dimension, the confinement problem depends only on the coordinate normal to the walls 
and the corresponding velocity component.

A particle initially located within the interval $x\in(0,L)$ eventually exits through one of the two walls.
The splitting probability is the probability that a particle initially located at $x$
reaches the wall at $x=L$ before reaching the wall at $x=0$.
For standard Brownian motion,
\begin{equation}
\pi_L(x)=\frac{x}{L},
\qquad \text{(diffusion)}
\label{eq:pi-diff}
\end{equation}
independently of the spatial dimension.
Consequently,
\[
\pi_L(0)=0,
\]
meaning that a particle initially located at the wall has zero probability of reaching the opposite boundary.

Introducing a constant drift does not alter this conclusion.
For Brownian motion with constant drift velocity $v$, the splitting probability depends on the initial position $x$, while
$v$ is a fixed parameter rather than a dynamical state variable.
To facilitate comparison with the active-particle models considered below, we denote it by $\pi_L(x;v)$.
It is given by
\begin{equation}
\pi_L(x;v)=
\frac{1-e^{-vx/D}}
     {1-e^{-vL/D}},
\qquad \text{(diffusion + drift)}
\label{eq:pi-drift}
\end{equation}
which again yields
\(
\pi_L(0;v)=0,
\)
for any $D>0$.
Thus, directed motion alone is insufficient to produce a finite wall splitting probability.
Only when $D=0$, that is, in the absence of thermal noise and when the motion becomes strictly ballistic, do we obtain
\(
\pi_L(0;v)=1,
\)
assuming $v>0$.

The microscopic origin of this result becomes evident from the short-time behavior of Brownian trajectories.
Consider a particle governed by
\be
dx = v\,dt + \sqrt{2D}\,dB_t,
\label{eq:dx-BM}
\ee
where $B_t$ denotes Brownian motion.
Since
\(
dB_t\sim dt^{1/2},
\)
whereas the deterministic displacement scales as \(dt\), stochastic fluctuations dominate in the limit \(dt\to0\).
Consequently, a Brownian trajectory starting at the wall crosses into the region \(x<0\) arbitrarily close to the initial time.
Strictly speaking, trajectories that reach the opposite wall without first crossing \(x=0\) do exist, but their probability is zero.
A more rigorous justification of this result, based on the classical theorem of Sparre Andersen, is provided in Appendix~\ref{app:SA}.

The situation changes qualitatively for active particles in the absence of translational diffusion.
Consider, for example, the active Ornstein--Uhlenbeck particle (AOUP),
\begin{align}
dx &= v(t)\,dt,
\nonumber\\
\tau_v\,dv &= -v(t)\,dt+\sqrt{2D_v}\,dB_t,
\label{eq:aoup}
\end{align}
where $v(t)$ is the swimming velocity,
$D_v$ is the diffusion coefficient governing fluctuations in the velocity,
and $\tau_v$ is the velocity relaxation time.
Unlike Brownian motion, stochastic fluctuations act on the velocity rather than directly on the particle position.
Consequently,
\begin{equation}
\Delta x=v(0)\Delta t+O(\Delta t^{3/2}),
\end{equation}
so that the particle position evolves ballistically at short times.
A particle initially located at the wall can therefore move away from the boundary before fluctuations significantly modify its swimming direction,
giving rise to
\[
\pi_L(0,v)>0,
\]
for $v>0$, where \(v\) denotes the component of the swimming velocity normal to the confining walls.

The same argument applies to the remaining canonical classes of active matter, including run-and-tumble particles (RTPs) and
active Brownian particles (ABPs).
In both cases the particle position evolves according to
\[
dx=v(t)\,dt,
\]
so that the short-time dynamics are likewise ballistic.
The condition $\pi_L(0)>0$, identified in Ref.~\cite{PRL-Klinger-2022} for discontinuous random walks,
is therefore not specific to jumping processes, but is a generic consequence of persistent motion in the absence of translational diffusion.

\section{Forward--backward formulation}
\label{sec:framework}

The canonical models considered here are Markovian only in the extended phase space $(x,v)$, 
where \(v\) denotes the component of the swimming velocity normal to the confining walls.  
Accordingly, all quantities introduced below are formulated in this state space.

The central quantities are the bulk stationary distribution $\rho(x,v)$ and the splitting probability $\pi_L(x,v)$.
Since a fraction of particles is dynamically adsorbed at the walls, the bulk distribution is normalized according to
\be
\int_0^L dx \int dv\,\rho(x,v) + 2f_w = 1,
\label{eq:rho-norm}
\ee
where $f_w$ denotes the fraction of particles adsorbed at each wall.  Throughout the remainder of this paper, we use the factorization
\be
\rho(x,v)=p_v(v)\,n(x,v),
\label{eq:factorization}
\ee
where $p_v(v)$ is the stationary marginal distribution of the velocity,
\[
p_v(v)=\int_0^L dx\,\rho(x,v)+f_0(v)+f_L(v), 
\]
with $f_0(v)$ and $f_L(v)$ denoting the velocity distributions of particles adsorbed at the left and right walls, respectively.
They satisfy
\[
f_w = \int dv\,f_0(v) = \int dv\,f_L(v),
\]
so that \(\int dv\,p_v(v) = 1\).  The explicit form of $p_v(v)$ depends on the active-particle model and is therefore left 
unspecified at this stage.  

The function $n(x,v)$ is the corresponding conditional bulk density.  
While the probability density $\rho(x,v)$ is defined over the full state space $(x,v)$, 
the conditional density $n(x,v)$ describes the distribution over position $x$ for a given velocity $v$.
Eq.~(\ref{eq:factorization}) is the standard decomposition of a joint probability density into its marginal and conditional 
distributions.

Because of the factorization, the two densities satisfy different forward equations.
The formulation based on $n(x,v)$ exhibits a closer correspondence with the backward equation for the splitting probability $\pi_L(x,v)$.
This makes the correspondence between the stationary density and the splitting probability more transparent.
Accordingly, the forward problem is formulated in terms of $n(x,v)$ throughout the remainder of this paper.

\subsection*{Dimensionless variables}

To place the three active-particle models within a common theoretical framework, we introduce the dimensionless variables
\[
(x,v)\rightarrow(z,w),
\qquad
z=\frac{x}{l_p},
\qquad
w=\frac{v}{v_{\rm char}},
\]
where $l_p$ is the persistence length and $v_{\rm char}$ is the characteristic velocity.  
Although the three models are governed by different microscopic parameters, each possesses natural characteristic 
scales of length and velocity.  These scales allow all three models to be expressed in a common dimensionless form.

For a run-and-tumble particle (RTP), the microscopic motion is governed by 
\begin{equation}
dx=v_0\cos\theta\,dt,
\end{equation}
where $\theta(t)$ denotes the angle between the swimming direction and the wall normal.  The orientation remains constant 
between successive tumbling events, whose waiting times are independently drawn from the exponential distribution
\[
p_t(t)=\tau^{-1}e^{-t/\tau},
\]
where $\tau$ is the mean waiting time.  After each tumbling event, a new orientation is drawn independently from the isotropic 
angular distribution.

For an active Brownian particle (ABP), the microscopic motion projected along the wall normal is governed by
\begin{align}
dx &= v_0\cos\theta\,dt,
\nonumber\\
d\theta &= (d-2)D_r\cot\theta\,dt+\sqrt{2D_r}\,dB_t,
\end{align}
The orientation evolves continuously through rotational diffusion with rotational diffusion coefficient $D_r$.
Finally, for an active Ornstein--Uhlenbeck particle (AOUP), the microscopic motion is governed by 
\begin{align}
dx &= v\,dt,
\nonumber\\
\tau_v\,dv &= -v\,dt+\sqrt{2D_v}\,dB_t.
\end{align}
The velocity evolves as an Ornstein--Uhlenbeck process with relaxation time $\tau_v$, while the parameter $D_v$ 
determines the strength of the velocity fluctuations.

The microscopic dynamics naturally define the characteristic velocity scale as
\[
v_{\rm char} =
\begin{cases}
v_0, & \text{RTP},\\
v_0, & \text{ABP},\\
\sqrt{D_v/\tau_v}, & \text{AOUP}.  
\end{cases}
\]
The corresponding persistence lengths are
\[
l_p=
\begin{cases}
v_{\rm char}\tau, & \mathrm{RTP},\\
v_{\rm char}/D_r, & \mathrm{ABP},\\
v_{\rm char}\tau_v, & \mathrm{AOUP},
\end{cases}
\]
The corresponding dimensionless system size is
\[
\lambda=\frac{L}{l_p}.
\]

\subsection{Forward equation}

Using the factorization \(\rho(z,w)=p_v(w)\,n(z,w)\) in dimensionless variables,
the stationary forward equation for $n(z,w)$ takes the form
\be
0 = -w\,\partial_z n(z,w) + \hat L_w n(z,w),
\label{eq:FS-v}
\ee
which is valid for all three active-particle models.
The spatial transport term $-w\,\partial_z n$ is universal and describes ballistic transport along the confinement direction.
The model dependence enters exclusively through the velocity operator $\hat L_w$, which governs the evolution of the projected velocity 
and reflects the underlying mechanism of active motion:
\begin{equation}
\hat L_w n=
\begin{cases}
-n+\int_{-1}^1dw'\,p_v(w')\,n(z,w'), & \text{RTP}, \\[2ex]
-(d-1)w\,\partial_w n + (1-w^2)\,\partial_w^2 n, & \text{ABP}, \\[2ex]
-w\,\partial_w n + \partial_w^2 n, & \text{AOUP}.
\end{cases}
\label{eq:L}
\end{equation}
The derivation of Eqs.~(\ref{eq:FS-v}) and (\ref{eq:L}) from the stationary forward equation
for the joint distribution $\rho(z,w)$ is given in Appendix~\ref{app:forward-rho}.

Besides reflecting different microscopic mechanisms, the velocity operators also exhibit different dependences on the spatial dimension $d$.
Only the AOUP operator is independent of the spatial dimension because each Cartesian velocity component evolves independently.
For ABP dynamics, the dimensional dependence appears explicitly through the drift term proportional to $(d-1)w$, whereas for RTP 
dynamics it enters implicitly through the projected velocity distribution $p_v(w)$.

The ABP and AOUP operators share a common drift--diffusion structure in the projected velocity $w$, whereas the RTP operator has 
a fundamentally different reaction-like structure.
To make the drift--diffusion structure explicit, the ABP operator can be rewritten as
\[
\hat L_w n =  -\partial_w \left[(d-3)wn\right]  +  \partial_w  \left[(1-w^2)\partial_w n\right]  +  (d-3)n
\]
from which one identifies
\[
D_{\rm eff}(w)=1-w^2
\]
as an effective diffusivity in projected-velocity space.
This diffusivity vanishes at $w=\pm1$, so changes in the projected velocity become increasingly slow near the extrema, reflecting the geometry of rotational diffusion.
For $d=3$, the drift term vanishes identically, leaving a particularly simple diffusion operator.

The RTP operator, by contrast, continuously removes particles from a given velocity state and redistributes them over the remaining velocity states through tumbling events.

\subsubsection*{Boundary conditions}

As discussed earlier, active motion in the absence of thermal fluctuations gives rise to dynamical adsorption at the confining walls.
The stationary state therefore comprises two populations: freely moving particles in the bulk and particles temporarily adsorbed at the walls.
The adsorbed particles are described by the wall distributions $f_0(w)$ and $f_\lambda(w)$, associated with the left and right walls, respectively.
These distributions enter the formulation through boundary conditions for the bulk distribution, which are determined by the exchange of particles 
between the bulk and the walls.
Since only particles whose velocity points toward a wall can remain adsorbed, the wall distributions have support only on incoming velocities: 
\begin{align}
f_0(w) &= 0, \qquad w>0,
\nonumber\\
f_\lambda(w) &= 0, \qquad w<0.
\label{eq:support-f}
\end{align}

Balancing the incoming and outgoing fluxes at each wall yields the boundary conditions
\begin{equation}
w\,n(0,w)=
\begin{cases}
f_w, & \text{RTP},\\
0, & \text{ABP and AOUP},
\end{cases}
\qquad w>0,
\label{eq:bc-left-n}
\end{equation}
and
\be
-\,w\,n(\lambda,w)=
\begin{cases}
f_w, & \text{RTP},\\
0, & \text{ABP and AOUP},
\end{cases}
\qquad w<0.
\label{eq:bc-right-n}
\ee
For RTP dynamics, reorientation is discrete, allowing particles to detach from the wall with a finite outward velocity and thereby 
producing a nonzero injection flux into the bulk.  For ABP and AOUP dynamics, reorientation is continuous, so particles leave 
the wall by passing through states with vanishing normal velocity, $w\approx0$.  Consequently, the injection flux vanishes for 
all finite outward velocities.  This distinction ultimately gives rise to the different wall singularities exhibited by the stationary 
distributions.

\subsection{Backward equation}

Let $\pi_\lambda(z,w)$ denote the probability that a particle initially located at position $z$ with velocity $w$ reaches the right boundary
$z=\lambda$ before the left boundary $z=0$.
It satisfies the backward equation
\be
0 = w\,\partial_z\pi_\lambda(z,w) + \hat L_w\pi_\lambda(z,w),
\label{eq:BS-w}
\ee
whose derivation is given in Appendix~\ref{app:BKE}.
The corresponding boundary conditions are
\begin{align}
\pi_\lambda(0,w)&=0, \qquad w<0,
\nonumber\\
\pi_\lambda(\lambda,w)&=1, \qquad w>0,
\label{eq:bc-pi}
\end{align}
reflecting the fact that a particle whose velocity points toward a wall is immediately absorbed.


Comparison of Eq.~(\ref{eq:BS-w}) with the forward equation~(\ref{eq:FS-v}) shows that the two equations differ only in the sign of the spatial transport term. 
This simple correspondence is made possible by the factorization
\(
\rho(z,w) = p_v(w)\,n(z,w),
\)
introduced earlier.

We also introduce the splitting probability at the wall, averaged over the stationary velocity distribution,
\be
\pi_\lambda(0)=\int dw\,p_v(w)\,\pi_\lambda(0,w).
\label{eq:pi-avg}
\ee
This quantity will play a central role in establishing the connection between the wall adsorption fraction $f_w$ and first-passage properties.

\subsection{Stochastic duality}

Comparing Eqs.~(\ref{eq:FS-v}) and (\ref{eq:BS-w}), we observe that the forward and backward equations differ only in the sign of the drift term.
We therefore consider the stationary distribution evaluated at the reversed velocity, \(n(z,-w)\).
Substituting \(w\rightarrow -w\) into the forward equation gives
\be
0 = w\,\partial_z n(z,-w)+\hat L_w n(z,-w).
\label{eq:dual-forward}
\ee
The forward equation satisfied by the velocity-reversed distribution therefore coincides with the backward equation for the original process.
This correspondence is an example of stochastic duality and, in the present setting, is closely related to Siegmund duality~\cite{AP-Siegmund-1976,JSTAT-Gueneau-2024}.
The dual process is obtained simply by reversing the swimming velocity, \(w\rightarrow -w\), so that particles move toward increasing \(z\) 
for \(w<0\) and toward decreasing \(z\) for \(w>0\).

The reversal of the velocity also transforms the boundary conditions.
Applying \(w\rightarrow -w\) to Eqs.~(\ref{eq:bc-left-n}) and (\ref{eq:bc-right-n}) yields
\begin{equation}
-\,w\, n(0,-w)=
\begin{cases}
f_w, & \text{RTP},\\
0, & \text{ABP and AOUP},
\end{cases}
\qquad w<0,
\label{eq:bc-left-aux}
\end{equation}
and
\be
w\, n(\lambda,-w)=
\begin{cases}
f_w, & \text{RTP},\\
0, & \text{ABP and AOUP},
\end{cases}
\qquad w>0.
\label{eq:bc-right-aux}
\ee
The transformation \(w\rightarrow -w\) preserves the total bulk probability,
\be
\int dw\,p_v(-w)\int_0^\lambda dz\,n(z,-w) = 1 - 2f_w.
\label{eq:ntilde-norm}
\ee
The corresponding conditional normalization becomes
\be
\int_0^\lambda dz\,n(z,-w) + \frac{f_0(-w) + f_\lambda(-w)}{p_v(-w)} = 1.
\label{eq:tilde-n-norm}
\ee
For the three canonical models considered here, the stationary projected velocity distribution is symmetric,
\(p_v(-w)=p_v(w)\).
Accordingly, Eq.~(\ref{eq:tilde-n-norm}) may be written with either $p_v(-w)$ or $p_v(w)$ in the denominator.

\subsection{Matching boundary conditions}

We have established that the velocity-reversed stationary distribution
$n(z,-w)$ and the splitting probability $\pi_\lambda(z,w)$ satisfy the same differential equation.
Their difference lies solely in the boundary conditions.
The remaining task is therefore to transform the two quantities in a way that leaves the differential equation 
unchanged while bringing their boundary conditions into exact agreement.

The first observation is that $n(z,-w)$ and $\pi_\lambda(z,w)$ exhibit different symmetries under reflection about the midpoint of the interval.
This difference can be eliminated by differentiating the splitting probability with respect to position.
The derivative $\partial_z\pi_\lambda(z,w)$ has the same reflection symmetry as $n(z,-w)$ and, since differentiation with respect to $z$ 
commutes with the differential operator, it satisfies the same differential equation.
It is therefore natural to seek a direct correspondence between $\partial_z\pi_\lambda(z,w)$ and $n(z,-w)$.

To determine the boundary condition for $\partial_z\pi_\lambda(z,w)$, we evaluate
Eq.~(\ref{eq:BS-w}) at $z=0$.
Together with the boundary condition~(\ref{eq:bc-pi}) for $w<0$, this yields
\begin{equation}
-w\,\partial_z\pi_\lambda(0,w)=
\begin{cases}
\pi_\lambda(0), & \text{RTP},\\[1ex]
0, & \text{ABP and AOUP},
\end{cases}
\qquad w<0,
\label{eq:bc-u-left}
\end{equation}
This boundary condition has the same structure as Eq.~(\ref{eq:bc-left-aux}), differing only by the multiplicative factor $\pi_\lambda(0)$ on the right-hand side.
Here, $\pi_\lambda(0)$ denotes the splitting probability evaluated at $z=0$ and averaged over the stationary velocity distribution, as defined in Eq.~(\ref{eq:pi-avg}).

Applying the same procedure at $z=\lambda$, together with the boundary condition
$\pi_\lambda(\lambda,w)=1$ for $w>0$, yields
\begin{equation}
w\,\partial_z\pi_\lambda(\lambda,w)=
\begin{cases}
\pi_\lambda(0), & \text{RTP},\\[1ex]
0, & \text{ABP and AOUP},
\end{cases}
\qquad w>0.
\label{eq:bc-u-right}
\end{equation}

We have therefore established that $\partial_z\pi_\lambda(z,w)$ and
$n(z,-w)$ satisfy the same differential equation and the same boundary
conditions up to a multiplicative constant.
By uniqueness of the corresponding boundary-value problem, the two
quantities must be proportional,
\be
\partial_z\pi_\lambda(z,w) = C\,n(z,-w).
\label{eq:u-prop-ntilde}
\ee

It remains to determine the proportionality constant $C$.
Integrating Eq.~(\ref{eq:u-prop-ntilde}) over $z$ and averaging over the stationary velocity distribution,
using Eq.~(\ref{eq:ntilde-norm}), gives
\be
C(1-2f_w) = 1 - 2\pi_\lambda(0).
\label{eq:C-normalization}
\ee 
For RTP, comparison of the boundary conditions for $\partial_z \pi_L(z,w)$ and $\tilde n(z,w)$ yields
\be
C = \frac{\pi_\lambda(0)}{f_w}.
\label{eq:C-RTP}
\ee
Combining Eqs.~(\ref{eq:C-RTP}) and (\ref{eq:C-normalization}) then gives
\(
C=1,
\)
and therefore
\[
\pi_\lambda(0) = f_w, \qquad \text{RTP}.
\]

For ABP and AOUP dynamics, the boundary conditions for $\partial_z\pi_\lambda(z,w)$ and $n(z,-w)$ are homogeneous.
Consequently, unlike the RTP case, the proportionality constant $C$ cannot be determined directly from the boundary conditions.
Instead, it is fixed by the exchange of particles between the wall and the bulk.
Integrating the forward equations for $n(z,-w)$ and $\partial_z\pi_\lambda(z,w)$ over the interval $z\in[0,\lambda]$ yields the corresponding wall equations.
For $w>0$ theses equations read 
\begin{align}
0 &= w\, n(0,-w) - w\,\partial_w \frac{f_\lambda(w)}{p_v(w)} + \partial_w^2 \frac{f_\lambda(w)}{p_v(w)}, \nonumber\\
0 &= w\, \partial_z \pi_L(0,w) - w\,\partial_w \pi_\lambda(0,w) + \partial_w^2 \pi_\lambda(0,w).
\label{eq:wall-match-continuous}
\end{align}

Multiplying the first equation by $C$ and using Eq.~(\ref{eq:u-prop-ntilde}) shows that $C f_\lambda(w)/p_v(w)$ and $\pi_\lambda(0,w)$ 
satisfy the same wall equation.  By uniqueness,
\be
p_v(w)\pi_\lambda(0,w) = C f_\lambda(w),
\qquad w>0.
\label{eq:wall-pi-f}
\ee

Integrating Eq.~(\ref{eq:wall-pi-f}) over $w>0$ yields
\(
C =  {\pi_\lambda(0)} / {f_w}.
\)
Combining this result with Eq.~(\ref{eq:C-normalization}) yields
\(
C=1,
\)
and therefore
\be
\pi_\lambda(0)=f_w,  \qquad \text{ABP and AOUP}.  
\ee

\section{Main identities}
\label{sec:results}

In the Introduction, we argued that dynamical adsorption and the condition
$\pi_L(0)>0$ originate from the same underlying mechanism, namely
directional persistence in the absence of translational diffusion.
The previous section established this correspondence through a rigorous
comparison of the forward and backward formulations.
The resulting exact identities are summarized below.

The rigorous correspondence established in the previous section culminates in the exact identity
\be
\partial_z\pi_\lambda(z,w)\equiv n(z,-w).
\label{eq:u-equals-ntilde}
\ee
All remaining identities follow directly from Eq.~(\ref{eq:u-equals-ntilde}).
Using the factorization
\(
\rho(z,w)=p_v(w)n(z,w),
\)
Eq.~(\ref{eq:u-equals-ntilde}) becomes
\be
p_v(w)\,\partial_z\pi_\lambda(z,w)=\rho(z,-w).
\label{eq:final-1}
\ee
Likewise, Eq.~(\ref{eq:wall-pi-f}) with $C=1$, or equivalently integrating Eq.~(\ref{eq:u-equals-ntilde}) over the interval \(z\in[0,\lambda]\), yields
\be
p_v(w)\,\pi_\lambda(0,w)=f_0(-w)=f_\lambda(w).
\label{eq:final-2}
\ee
Finally, defining the velocity-averaged splitting probability as
\(
\pi_\lambda(z)=\int dw\,p_v(w)\,\pi_\lambda(z,w),
\)
integration of Eqs.~(\ref{eq:final-1}) and (\ref{eq:final-2}) over the velocity
yields the corresponding marginal identities
\begin{align}
\partial_z\pi_\lambda(z) &= \rho(z), \nonumber\\
\pi_\lambda(0) &= f_w.
\label{eq:final-marginal}
\end{align}
The second identity establishes that the fraction of dynamically adsorbed particles at a wall, \(f_w\), 
is exactly equal to the probability that a particle starting from that wall reaches the opposite wall before returning.

The main identity in Eq. (\ref{eq:u-equals-ntilde}), and all the identities derived from it,
are valid for all three models of active dynamics, RTP, ABP, and AOUP, and in arbitrary spatial dimension $d$.  
The last identity establishes the precise relation that originally motivated this work: the
probability of reaching the opposite wall when starting from a confining wall
is encoded exactly in the stationary fraction of particles adsorbed at that
wall. More generally, the correspondence reveals a hierarchy of exact
identities linking stationary and first-passage observables.

The identities derived above are completely model independent. 
Their only model-dependent ingredient is the stationary velocity distribution \(p_v(w)\), 
which for the three canonical active-particle models is given by
\be
p_v(w)=
\begin{cases}
\displaystyle
\frac{\Gamma(d/2)}{\sqrt{\pi}\,\Gamma(d/2-1/2)}
(1-w^2)^{(d-3)/2},
& \text{RTP and ABP},
\\[2ex]
\displaystyle
\frac{e^{-w^2/2}}{\sqrt{2\pi}},
& \text{AOUP}.
\end{cases}
\label{eq:pv-all}
\ee
In all cases, the stationary velocity distribution is symmetric,
\[
p_v(w)=p_v(-w).
\]

\subsection{Stochastic duality}

This work was motivated by the hypothesis that dynamical adsorption and the condition $\pi_L(0)>0$ are intimately connected.
To establish this connection rigorously, we compared the forward and backward formulations governing these two phenomena.
This comparison was carried out entirely at the level of their mathematical structure. 
As a result, the physical interpretation of the resulting duality is not immediately apparent

In retrospect, the correspondence established here belongs to the broader class of Siegmund dualities.  
More specifically, it belongs to the class of Siegmund dualities
introduced by Siegmund~\cite{AP-Siegmund-1976} and recently discussed in the
context of active particles by Guéneau and Touzo~\cite{JSTAT-Gueneau-2024}.

The identities derived in this work also provide a new physical interpretation of the splitting probability. The relation
\[
\partial_z\pi_\lambda(z,w)=n(z,-w)
\]
shows that the splitting probability increases as the particle moves toward the wall at \(z=\lambda\), since \(n(z,-w)>0\).
Although this monotonic increase is expected, its local rate is determined not by the stationary distribution of the original process, 
but by the stationary distribution of its dual, velocity-reversed process.

to state it differently, an observable of one process is related to a different observable of another process. 
Neither the observables nor the processes can be arbitrary.   The microscopically,

\section{Generalized RTP model}

Before verifying the relations derived in this work, we consider a generalized RTP model. This extension is motivated by the 
fact that an RTP can be interpreted as a jumping-particle process by recording the particle position only at tumble events 
\cite{POF-Frydel-2024}. For exponentially distributed run times, this jumping representation yields the same stationary distribution 
as the original continuous RTP dynamics.

We therefore relax the standard RTP assumption of a fixed swimming speed and allow an arbitrary symmetric, normalized velocity 
distribution,
\[
p_v(w)=p_v(-w),
\qquad
\int dw\,p_v(w)=1,
\]
following Ref.~\cite{JSTAT-Frydel-2021}.


The jumping-particle model is characterized by its jump-length distribution,
denoted by \(G(\ell)\).
Since the generalized RTP model is specified through the velocity distribution
\(p_v(v)\), the corresponding jump-length distribution is obtained from
\begin{equation}
G(\ell)=
\int_0^\infty dt\,p_t(t)
\int_{-\infty}^{\infty}dv\,p_v(v)\,
\delta(\ell-vt),
\label{eq:G}
\end{equation}
where the Dirac delta,
\(
\delta(\ell-vt),
\)
is the propagator of a particle moving ballistically with constant velocity \(v\).
Since the velocity and run time are sampled independently from the distributions
\(p_v(v)\) and \(p_t(t)\), respectively, the jump-length distribution is obtained by averaging the ballistic propagator over all possible realizations.

For the standard RTP model, the run-time distribution is exponential,
\[
p_t(t)=\tau^{-1}e^{-t/\tau},
\]
which renders the RTP dynamics Markovian. Performing the velocity integral gives
\begin{equation}
G(\ell)
=
\frac{1}{\tau}
\int_0^\infty dt\,
e^{-t/\tau}
\frac{1}{t}
p_v\!\left(\frac{\ell}{t}\right).
\label{eq:G-exp}
\end{equation}
The resulting jump-length distribution therefore depends on the particular choice of \(p_v(v)\).

To establish a connection with the work of Klinger \textit{et al.}~\cite{PRL-Klinger-2022},
we consider a velocity distribution with an algebraic tail,
\be
p_v(v)\sim A |v|^{-\alpha},
\qquad |v|\to\infty,
\ee
with \(\alpha>1\).
Substituting this form into Eq.~(\ref{eq:G-exp}) gives
\be
G(\ell)
\sim
A\,\Gamma(\alpha)\,
\tau^{\alpha-1}
|\ell|^{-\alpha},
\qquad |\ell|\to\infty.
\ee
Thus, an algebraic velocity distribution generates a jump-length distribution with the same algebraic tail.

To establish a connection with the work of Klinger \textit{et al.}~\cite{PRL-Klinger-2022},
we consider a velocity distribution with an algebraic tail,
\begin{equation}
p_v(v)\sim A|v|^{-\alpha},
\qquad |v|\to\infty,
\end{equation}
with \(\alpha>1\).
Substituting this form into Eq.~(\ref{eq:G-exp}) gives
\begin{equation}
G(\ell)
\sim
A\,\Gamma(\alpha)\,
\tau^{\alpha-1}
|\ell|^{-\alpha},
\qquad |\ell|\to\infty.
\end{equation}
Thus, an algebraic velocity distribution generates a jump-length distribution with the same algebraic tail.

Klinger \textit{et al.}~\cite{PRL-Klinger-2022} obtained the asymptotic splitting probability for jump processes in slab geometry.
For \(\alpha>3\), the jump-length distribution has finite variance and
\begin{equation}
\pi_L(0)\propto L^{-1}.
\end{equation}
For \(1<\alpha<3\), the jump-length distribution exhibits Lévy-flight statistics with divergent variance, leading to
\be
\pi_L(0)\propto L^{-\frac{\alpha}{2}+\frac12}.
\label{eq:pi-tail}
\ee
Using the correspondence
\(
\pi_L(0)=f_w,
\)
derived in the present work, we immediately obtain
\be
f_w(L)\propto L^{-\frac{\alpha}{2}+\frac12},
\qquad 1<\alpha<3.
\label{eq:fw-tail}
\ee
Thus, the first-passage results of Ref.~\cite{PRL-Klinger-2022} immediately yield exact predictions for the stationary 
fraction of particles dynamically adsorbed at the confining walls.

\section{Numerical verification}
\label{sec:numerics}

In this section, we verify the identities derived in this work by numerical simulations of RTP, ABP, and AOUP particles 
at zero temperature confined between two parallel hard walls.  All results presented below correspond to three spatial dimensions (\(d=3\)).

The first prediction of the theory is the identity
\(
\pi_\lambda(0)=f_w.
\)
To test this relation, the two quantities are obtained from independent numerical simulations.
The fraction of particles adsorbed at a wall, \(f_w\), is obtained by evolving the particle for a sufficiently long time and sampling its position at regular time intervals.
It is then given by the fraction of sampled configurations in which the particle is adsorbed at a given wall.
The splitting probability \(\pi_\lambda(0)\), by contrast, is obtained from independent first-passage simulations in which particles are initialized at the left wall.
It is given by the fraction of trajectories that reach the opposite wall before returning to the initial wall.

\begin{figure}[t]
\centering
\includegraphics[width=0.6\linewidth]{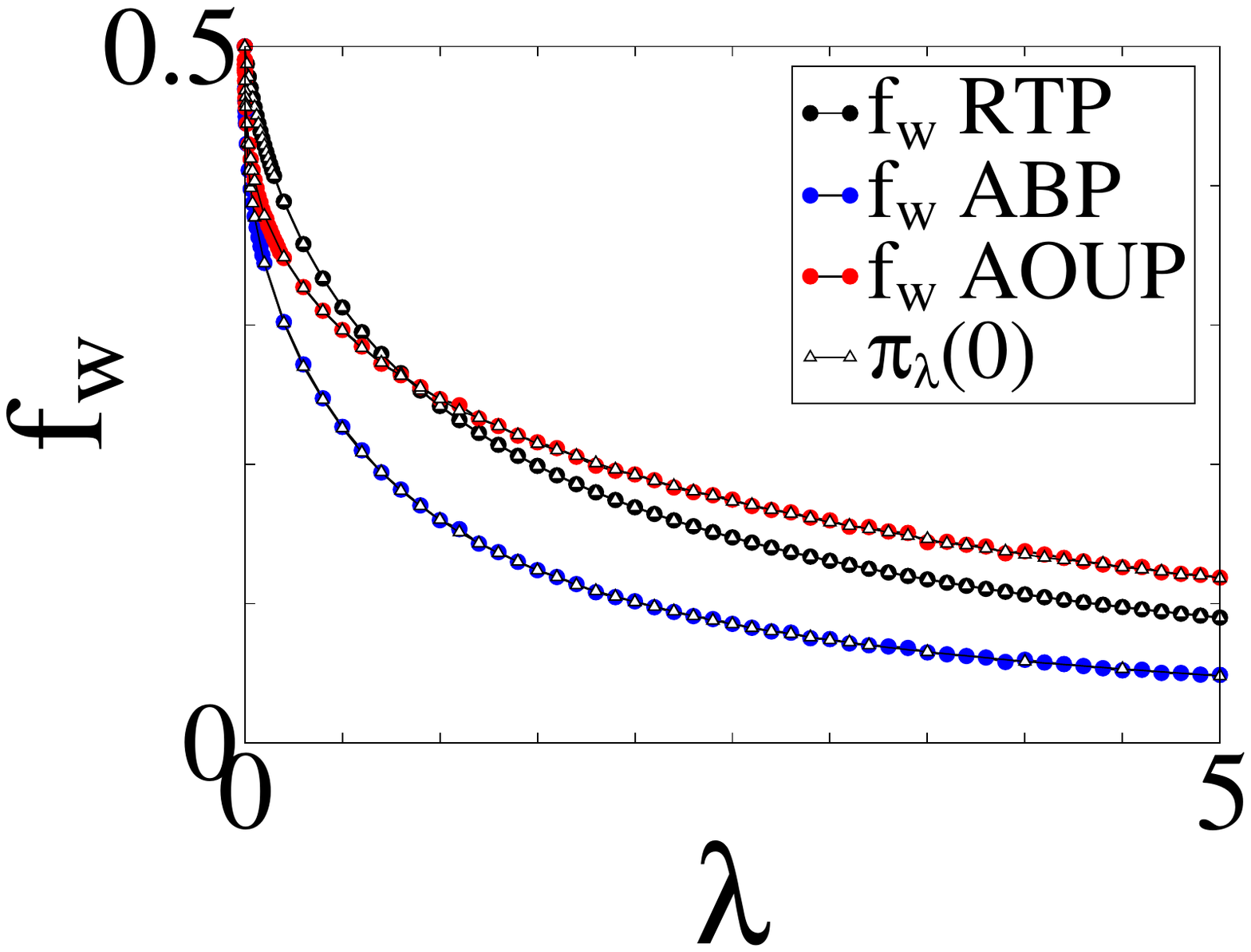}
\caption{
Fraction of particles adsorbed at one of the confining walls, \(f_w\), and the splitting probability
\(\pi_\lambda(0)\) as functions of the wall separation \(\lambda\) for RTP, ABP, and AOUP dynamics in three dimensions.
The two quantities are indistinguishable within statistical error.
}
\label{fig:fw-pi}
\end{figure}

Fig.~\ref{fig:fw-pi} compares the two quantities for RTP, ABP, and AOUP dynamics.
The agreement within statistical error confirms the theoretical prediction.
In all three models, the fraction of particles adsorbed at a wall and the splitting probability exhibit the common asymptotic behavior
\[
\pi_\lambda(0)=f_w(\lambda)\propto\lambda^{-1},
\qquad \lambda\gg1.
\]
This scaling reflects the crossover to effective Brownian diffusion on length scales much larger than the persistence length.
In this regime, the splitting probability reduces to the Brownian result given by Eq.~(\ref{eq:pi-diff}).

\begin{figure}[t]
\centering

\begin{minipage}{0.45\linewidth}
\centering
\includegraphics[width=\linewidth]{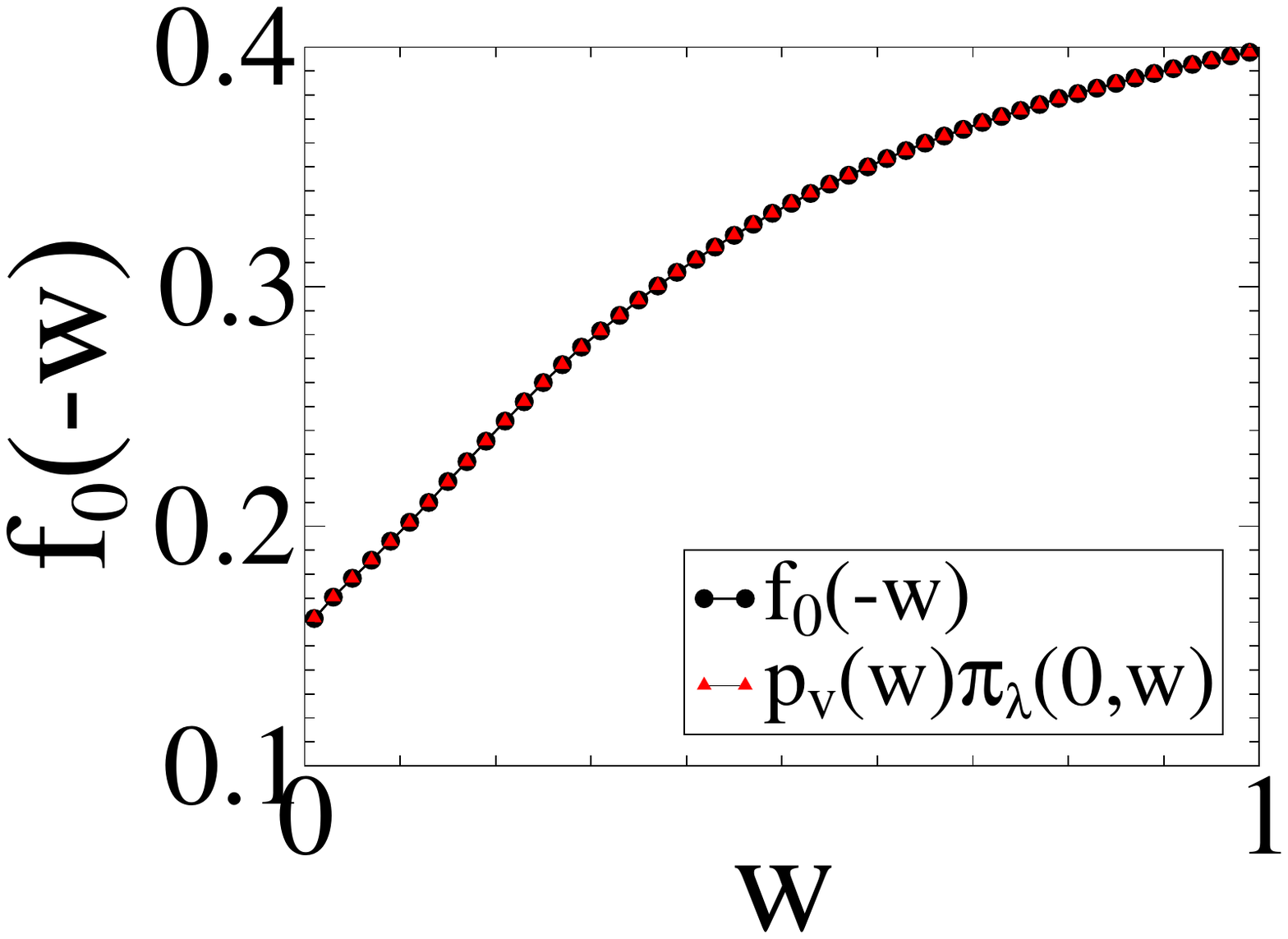}
\end{minipage}

\vspace{0.3cm}

\begin{minipage}{0.45\linewidth}
\centering
\includegraphics[width=\linewidth]{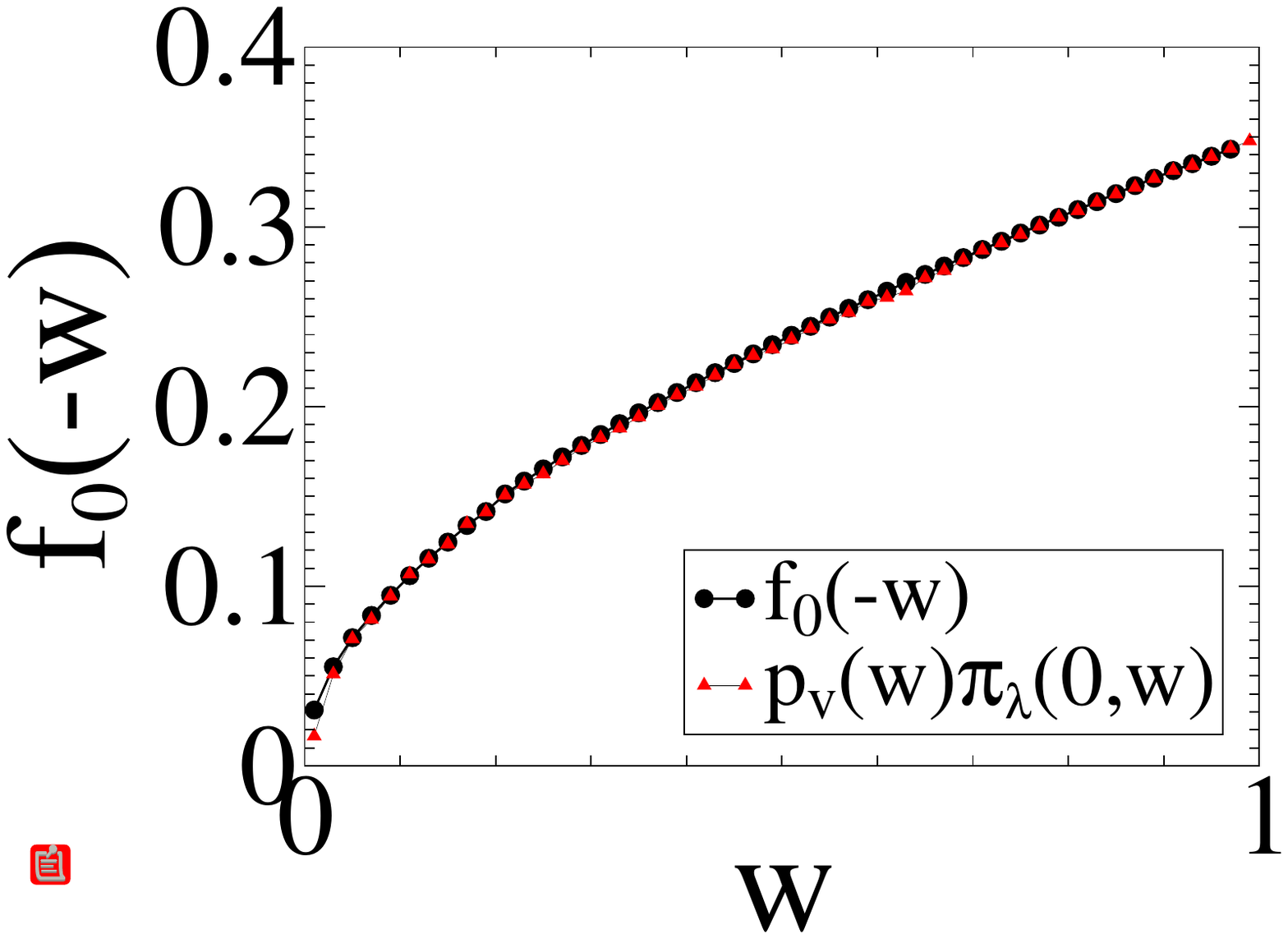}
\end{minipage}

\vspace{0.3cm}

\begin{minipage}{0.45\linewidth}
\centering
\includegraphics[width=\linewidth]{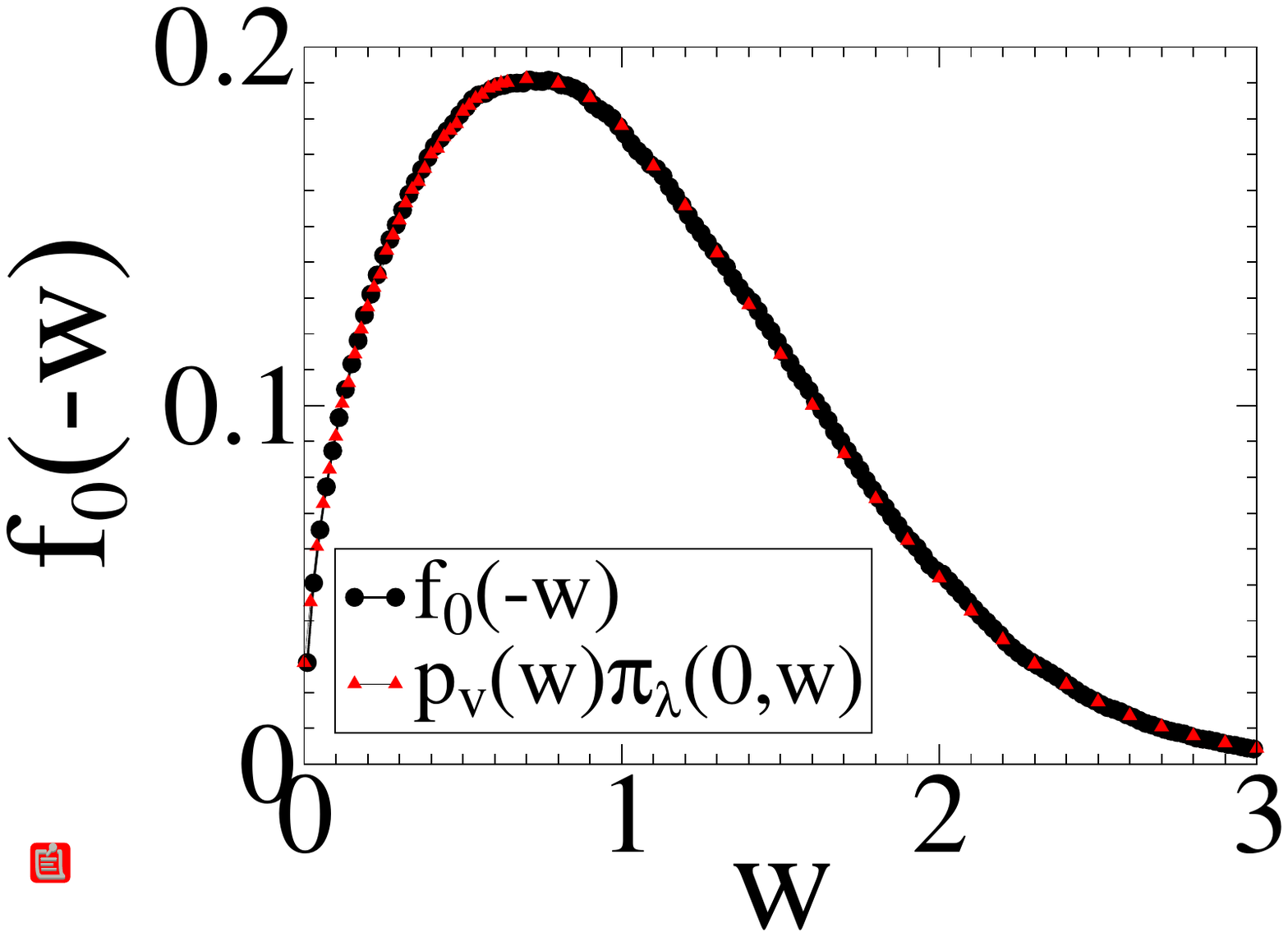}
\end{minipage}

\caption{
Velocity-resolved comparison between \(p_v(w)\pi_\lambda(0,w)\) and \(f_0(-w)\) for RTP, ABP, and AOUP dynamics in three spatial dimensions.
}
\label{fig:f0-three}
\end{figure}

We next test the stronger velocity-resolved identity
\[
p_v(w)\pi_\lambda(0,w)=f_0(-w).
\]
Fig.~\ref{fig:f0-three} compares
\(p_v(w)\pi_\lambda(0,w)\)
and
\(f_0(-w)\)
for RTP, ABP, and AOUP dynamics in three spatial dimensions at
\(\lambda=1/2\).
The two quantities are indistinguishable within statistical error, confirming the velocity-resolved identity.

\begin{figure}[t]
\centering
\includegraphics[width=0.6\linewidth]{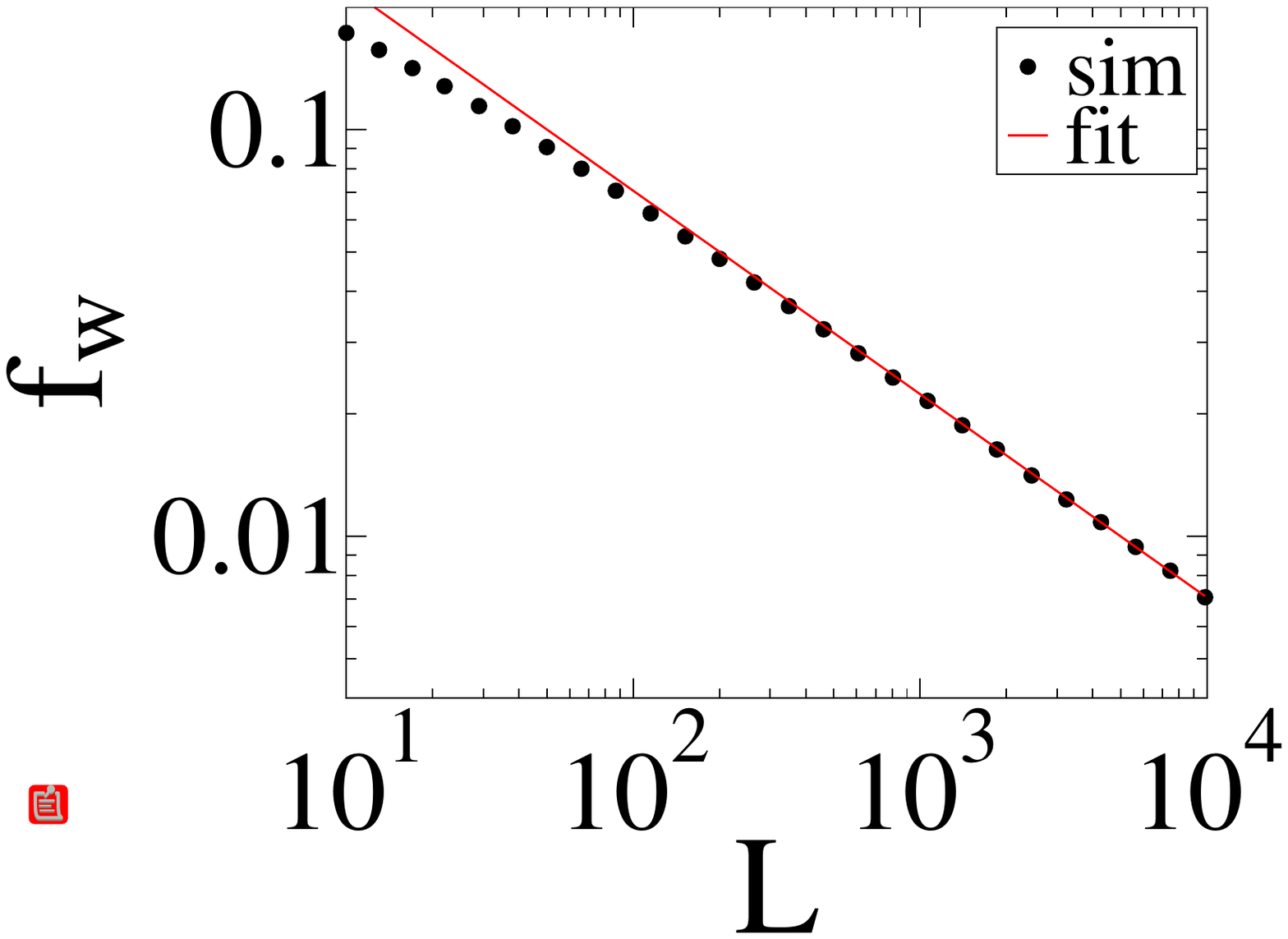}
\caption{
Fraction of particles adsorbed at a confining wall, \(f_w\), for a generalized RTP model with
\(p_v(v)\sim |v|^{-2}\) and exponentially distributed run times \((\tau=1)\).
The dashed line shows the asymptotic prediction \(f_w\propto L^{-1/2}\).
}
\label{fig:fw-a2}
\end{figure}

As a final application, we demonstrate how the correspondence can be used to convert known first-passage results into predictions for stationary adsorption.
We consider a generalized RTP model with an algebraic velocity distribution,
\(
p_v(v)\sim |v|^{-2}.
\)
According to Eq.~(\ref{eq:pi-tail}), the splitting probability obeys the asymptotic scaling
\[
\pi_L(0)\propto L^{-1/2}.
\]
Using the identity
\(
\pi_L(0)=f_w,
\)
derived in the present work, we immediately predict
\[
f_w\propto L^{-1/2}.
\]
Fig.~\ref{fig:fw-a2} confirms the predicted asymptotic scaling for the stationary wall fraction.

\section{Conclusion}

This work begins with the observation that, unlike standard Brownian motion, active particles in the absence 
of thermal fluctuations exhibit a finite wall splitting probability,
\[
\pi_L(0)>0,
\]
meaning that a particle initially located at one confining wall has a nonzero probability of reaching the opposite wall before 
returning to its initial boundary.  At the same time, such systems exhibit dynamical adsorption at the confining walls.  
The coexistence of these two phenomena naturally raises the question of whether they originate from the same physical 
mechanism and whether they are quantitatively related.

To answer this question, we formulated stationary distributions and splitting probabilities within a common position--velocity 
phase-space framework.  By comparing the corresponding forward and backward operators, we established an exact 
correspondence between stationary and first-passage observables for the three canonical classes of active dynamics: RTPs, ABPs, and AOUPs.

The analysis led to the exact relations
\begin{align}
p_v(w)\,\partial_z \pi_\lambda(z,w) &= \rho(z,-w), \nonumber\\
p_v(w)\,\pi_\lambda(0,w) &= f_0(-w),
\end{align}
and, after averaging over the velocity distribution,
\begin{align}
\partial_z \pi_\lambda(z) &= \rho(z), \nonumber\\
\pi_\lambda(0) &= f_w.
\end{align}

These identities establish an exact bridge between stationary and first-passage descriptions of confined active matter.
In particular, they show that the fraction of particles dynamically adsorbed at a confining wall is exactly equal to the probability 
that a particle starting from that wall reaches the opposite boundary.  More fundamentally, they demonstrate that dynamical 
adsorption and finite wall splitting probabilities are complementary manifestations of the same persistence mechanism.

Finally, we showed that the framework extends naturally to generalized RTPs with arbitrary velocity distributions, thereby establishing 
a direct connection with jumping-particle processes.  This extension allows known first-passage results for jump processes to be translated 
directly into exact predictions for stationary adsorption, illustrating that the correspondence extends beyond the canonical active-particle 
models considered here.

\vspace{0.5cm}
\begin{acknowledgments}

D.F. acknowledges financial support from FONDECYT through grant number 1241694.  

\end{acknowledgments}

\section{DATA AVAILABILITY}
The data that support the findings of this study are available from the corresponding author upon 
reasonable request.

\appendix

\section{Sparre Andersen theorem}
\label{app:SA}

In Sec.~\ref{sec:trajectories}, immediately after Eq.~(\ref{eq:dx-BM}), we argued that a Brownian particle initially located at a confining wall satisfies
\(
\pi_L(0)=0,
\)
because Brownian fluctuations dominate the deterministic drift arbitrarily close to the initial time.
Here we provide a more rigorous justification based on the classical theorem of Sparre Andersen~\cite{SparreAndersen1953,SparreAndersen1954}.

Consider a symmetric random walk starting at the origin.  Let
\(
S_n
\)
denote the probability that the trajectory never becomes negative during the first \(n\) steps.  Sparre Andersen's theorem states that
\be
S_n = \frac{1}{2^{2n}} \frac{(2n)!}{(n!)^2} = \frac{1}{\sqrt{\pi}} \frac{\Gamma\!\left(n+\frac12\right)} {\Gamma(n+1)}.
\label{eq:SA}
\ee
This expression is universal within the class of symmetric random walks and depends only on the number of steps \(n\).
Remarkably, it does not depend on the particular distribution of jumps.
For Brownian motion, these increments are Gaussian distributed.

To relate this result to Brownian motion, we discretize a fixed time interval \(t\) into
\[
n = \frac{t}{\Delta t}
\]
independent Brownian increments. Such a discretization is routinely employed in Brownian dynamics simulations.
Using the asymptotic form of Eq.~(\ref{eq:SA}) for large \(n\),
\(
S_n \sim \frac{1}{\sqrt{\pi n}},
\)
we obtain
\[
S_n \sim \sqrt{\frac{\Delta t}{\pi t}}.
\]
Therefore, for any fixed observation time \(t>0\),
\[
\lim_{\Delta t\rightarrow0}S_n = 0.
\]
In the context of numerical simulations, a finite integration time step \(\Delta t\) can occasionally produce trajectories that reach 
the opposite wall without first crossing \(x=0\).  However, as the time step is refined, the probability of such trajectories decreases as
\[
S_n\sim\sqrt{\Delta t},
\]
and therefore vanishes in the continuous-time limit.


Thus, the probability that a Brownian trajectory starting at the wall remains entirely in the region \(x\ge0\)
during any finite time interval vanishes. Since reaching the opposite wall without first entering \(x<0\)
requires precisely such trajectories, it follows that
\[
\pi_L(0)=0.
\]
The result demonstrates that the vanishing wall splitting probability is a direct consequence of the microscopic roughness of Brownian trajectories.
Therefore, any finite value of
\(
\pi_L(0)>0
\)
must originate from ballistic short-time dynamics.

\section{Derivation of the forward equation for the conditional density}
\label{app:forward-rho}

In the main text, the stationary forward problem is formulated in terms of the conditional density
\(
\rho(z,w)=p_v(w)\,n(z,w),
\)
where $p_v(w)$ is the stationary marginal distribution of the projected velocity.
For completeness, we briefly show how the corresponding forward equation for $n(z,w)$ follows from the stationary equation 
for the joint distribution $\rho(z,w)$.

For all three active-particle models, the stationary forward equation has the general form
\begin{equation}
0=-w\,\partial_z\rho(z,w)+\hat O_w\rho(z,w),
\label{eq:rho-forward}
\end{equation}
where the transport term is universal, while the velocity operator $\hat O_w$ depends on the microscopic dynamics:
\be
\hat O_w\rho =
\begin{cases}
-\rho +p_v(w)\int_{-1}^{1}dw'\,\rho(z,w'), & \text{RTP}, \\[2ex]
(d-1)\partial_w(w\rho) + \partial_w^2\!\left[(1-w^2)\rho\right], & \text{ABP}, \\[2ex]
\partial_w(w\rho) + \partial_w^2\rho, & \text{AOUP}.
\end{cases}
\label{eq:Lrho}
\end{equation}
The stationary velocity distribution satisfies
\begin{equation}
\hat O_w p_v(w)=0.
\label{eq:pv-stationary}
\ee
Explicitly,
\be
0 =
\begin{cases}
-p_v(w) + p_v(w)\int_{-1}^{1}dw'\,p_v(w'), & \text{RTP}, \\[2ex]
(d-1)\partial_w\!\left[w\,p_v(w)\right] + \partial_w^2\!\left[(1-w^2)p_v(w)\right], & \text{ABP}, \\[2ex]
\partial_w\!\left[w\,p_v(w)\right] + \partial_w^2p_v(w), & \text{AOUP}.
\end{cases}
\label{eq:pv-equations}
\ee
Note that the RTP equation is satisfied by any normalized distribution,
\(
\int_{-1}^1dw\,p_v(w)=1.
\)
Consequently, unlike the ABP and AOUP cases, the stationary projected-velocity
distribution is not determined by Eq.~(\ref{eq:pv-equations}) itself, but by the
underlying isotropic distribution of swimming directions.

Substituting
\(
\rho(z,w)=p_v(w)\,n(z,w)
\)
into Eq.~(\ref{eq:rho-forward}) and using Eq.~(\ref{eq:pv-equations}), all terms involving the stationary velocity distribution combine to give
\begin{equation}
\hat O_w\!\left[p_v(w)n(z,w)\right] = p_v(w)\,\hat L_wn(z,w),
\label{eq:operator-factorization}
\end{equation}
where the operator $\hat L_w$ is given in Eq.~(\ref{eq:L}).
Dividing Eq.~(\ref{eq:rho-forward}) by $p_v(w)$ therefore yields
\be
0= - w\,\partial_zn(z,w) + \hat L_wn(z,w),
\ee
which is Eq.~(\ref{eq:FS-v}) of the main text.

\section{Derivation of the backward equation}
\label{app:BKE}

In this appendix, we briefly derive the backward equation satisfied by the splitting probability.
Consider a particle initially prepared in the state \((z,w)\).  The splitting probability, \(\pi_\lambda(z,w)\), is the probability 
that the particle reaches the wall at \(z=\lambda\) before returning to the wall at \(z=0\).
Suppose the particle evolves over an infinitesimal time interval \(ds\),
\[
(z,w)\rightarrow(z+dz,w+dw),
\]
where the increments \(dz\) and \(dw\) are random variables determined by the stochastic dynamics. We further introduce the dimensionless time
\[
s=\frac{t}{\tau_p},
\qquad
\tau_p=\frac{l_p}{v_{\rm char}},
\]
where \(\tau_p\) denotes the relevant persistence time.

The trajectory then branches into infinitely many possible realizations.
The splitting probability associated with the initial state must therefore be preserved upon averaging over all such realizations.
Consequently,
\be
\pi_\lambda(z,w) = \left\langle \pi_\lambda(z+dz,w+dw) \right\rangle,
\label{eq:backward-average}
\ee
where the average is taken over all realizations of the stochastic increments.

Expanding the right-hand side of Eq.~(\ref{eq:backward-average}) gives
\begin{align}
\pi_\lambda(z+dz,w+dw) &=  \pi_\lambda(z,w) + \partial_z\pi_\lambda\,dz + \partial_w\pi_\lambda\,dw \nonumber\\ 
& + \frac12\partial_z^2\pi_\lambda\,(dz)^2 + \frac12\partial_w^2\pi_\lambda\,(dw)^2 \nonumber\\
& + \partial_{zw}\pi_\lambda\,dz\,dw  + \cdots
\end{align}
Inserting this in Eq. (\ref{eq:backward-average}) and using $dz = w ds$ and keeping terms to first order yields 
\begin{align}
0&  =  w\partial_z\pi_\lambda\, ds   +   \partial_w\pi_\lambda\,\langle dw\rangle   +   \frac12\partial_w^2\pi_\lambda\, \langle (dw)^2\rangle 
\label{eq:backward-average-2}
\end{align}

For active Brownian and active Ornstein--Uhlenbeck particles,
\ba
&& dw = -(d-1)w\,ds + \sqrt{2(1-w^2)}\, dB_s, \qquad \hspace{-0.2cm} \text{ABP}  \nonumber\\
&& dw = -w\,ds + \sqrt{2}\,dB_s, \qquad\qquad\qquad\qquad \text{AOUP}  \nonumber
\ea
so that
\ba
&& \langle dw\rangle=-(d-1)w\,ds, \qquad \langle(dw)^2\rangle = 2(1-w^2)\,ds, \nonumber\\
&& \langle dw\rangle = -w\,ds, \qquad\qquad \quad  \langle(dw)^2\rangle = 2ds,  \nonumber
\ea
Substituting these moments into Eq.~(\ref{eq:backward-average-2}) gives
\ba
&& 0 = w\partial_z\pi_\lambda - w(d-1)\partial_w\pi_\lambda + (1-w^2)\partial_w^2 \pi_\lambda, \nonumber\\
&& 0 = w\partial_z\pi_\lambda - w\partial_w\pi_\lambda + \partial_w^2\pi_\lambda, \nonumber
\ea
for ABP and AOUP, respectively. The result is precisely the backward equation used in the main text
and given as $0 = w\partial_z\pi_\lambda  +  \hat L_w \pi_\lambda$ in Eq. (\ref{eq:BS-w}).  
The corresponding equation for RTPs follows analogously from the first two moments of their respective stochastic increments.

The RTP dynamics require a separate treatment because the velocity undergoes finite jumps rather than infinitesimal increments.
and the Taylor expansion in \(dw\) used above is not applicable. In the dimensionless time variable \(s\), the tumbling
rate is unity. Thus, during an infinitesimal interval \(ds\), the particle does not tumble with probability \(1-ds\), while it tumbles
with probability \(ds\), up to terms of order \(ds^2\). If no tumble occurs, the particle moves ballistically with velocity \(w\).
If a tumble occurs, a new velocity \(w'\) is drawn from the distribution \(p_v(w')\).
Consequently, to first order in \(ds\),
\ba
\pi_\lambda(z,w) &=& (1-ds)\pi_\lambda(z+w\,ds,w) \nonumber\\
&+& ds\int dw'\,p_v(w')\,\pi_\lambda(z+w\,ds,w').
\label{eq:backward-RTP}
\ea
Using the expansion 
\[
\pi_\lambda(z+w\,ds,w) = \pi_\lambda(z,w) + w\,ds\,\partial_z\pi_\lambda(z,w) + \cdots,
\]
we obtain to first order 
\be
0 = w\,\partial_z\pi_\lambda(z,w) - \pi_\lambda(z,w) + \int dw'\,p_v(w')\,\pi_\lambda(z,w').
\ee
This is precisely the RTP backward equation given in Eq.~(\ref{eq:BS-w}).

\bibliography{general-fixed}

@article{AP-Siegmund-1976,
  author    = {Siegmund, D.},
  title     = {The Equivalence of Absorbing and Reflecting Barrier Problems for Stochastically Monotone {M}arkov Processes},
  journal   = {The Annals of Probability},
  volume    = {4},
  number    = {6},
  pages     = {914--924},
  month     = {December},
  year      = {1976},
  publisher = {Institute of Mathematical Statistics},
  doi       = {10.1214/aop/1176995936},
  url       = {https://doi.org/10.1214/aop/1176995936}
}

@article{JPA-Gueneau-2024,
doi = {10.1088/1751-8121/ad4753},
url = {https://doi.org/10.1088/1751-8121/ad4753},
year = {2024},
month = {may},
publisher = {IOP Publishing},
volume = {57},
number = {22},
pages = {225005},
author = {Guéneau, Mathis and Touzo, Léo},
title = {Relating absorbing and hard wall boundary conditions for a one-dimensional run-and-tumble particle},
journal = {Journal of Physics A: Mathematical and Theoretical}
}

@article{JSTAT-Gueneau-2024,
doi = {10.1088/1742-5468/ad6134},
url = {https://doi.org/10.1088/1742-5468/ad6134},
year = {2024},
month = {sep},
publisher = {IOP Publishing},
volume = {2024},
number = {8},
pages = {083208},
author = {Guéneau, Mathis and Touzo, Léo},
title = {Siegmund duality for physicists: a bridge between spatial and first-passage properties of continuous- and discrete-time stochastic processes},
journal = {Journal of Statistical Mechanics: Theory and Experiment}
}

@article{Baouche-2026-archive,
  author        = {Baouche, Y. and Gu{\'e}neau, M. and Kurzthaler, C.},
  title         = {Spatiotemporal characterization of active {B}rownian dynamics in channels},
  journal       = {arXiv preprint arXiv:2603.12080},
  year          = {2026},
  eprint        = {2603.12080},
  archivePrefix = {arXiv},
  primaryClass  = {physics.bio-ph},
  url           = {https://arxiv.org/abs/2603.12080}
}

@article{AP-Bray-2013,
  author    = {Bray, A. J. and Majumdar, S. N. and Schehr, G.},
  title     = {Persistence and first-passage properties in nonequilibrium systems},
  journal   = {Advances in Physics},
  volume    = {62},
  number    = {3},
  pages     = {225--361},
  year      = {2013},
  publisher = {Taylor \& Francis},
  doi       = {10.1080/00018732.2013.803819},
  url       = {https://doi.org/10.1080/00018732.2013.803819}
}

@Inbook{book-Basu-2024,
author="Basu, Urna
and Sabhapandit, Sanjib
and Santra, Ion",
editor="Grebenkov, Denis
and Metzler, Ralf
and Oshanin, Gleb",
title="Target Search by Active Particles",
bookTitle="Target Search Problems",
year="2024",
publisher="Springer Nature Switzerland",
address="Cham",
pages="463--487",
isbn="978-3-031-67802-8",
doi="10.1007/978-3-031-67802-8_19",
url="https://doi.org/10.1007/978-3-031-67802-8_19"
}

@article{Oshanin-2012,
  title = {First-passage phenomena and their applications},
  author = {Oshanin, Gleb and Metzler, Ralf and Schehr, Gregory},
  journal = {World Scientific},
  year = {2012},
}

@article{Demaerel2018,
  title = {Active processes in one dimension},
  author = {Demaerel, Thibaut and Maes, Christian},
  journal = {Phys. Rev. E},
  volume = {97},
  issue = {3},
  pages = {032604},
  numpages = {12},
  year = {2018},
  month = {Mar},
  publisher = {American Physical Society},
  doi = {10.1103/PhysRevE.97.032604},
  url = {https://link.aps.org/doi/10.1103/PhysRevE.97.032604},
  note = {Rigorous treatment of backward/forward operators in active 1D systems.}
}

@article{Basu2019,
  title = {Active Brownian motion in two dimensions},
  author = {Basu, U. and Majumdar, S. N. and Rosso, A. and Schehr, G.},
  journal = {Phys. Rev. E},
  volume = {100},
  pages = {062116},
  year = {2019}
}

@article{PRL-Mori-2020,
  title = {Universal Survival Probability for a $d$-Dimensional Run-and-Tumble Particle},
  author = {Mori, Francesco and Le Doussal, Pierre and Majumdar, Satya N. and Schehr, Gr\'egory},
  journal = {Phys. Rev. Lett.},
  volume = {124},
  pages = {090603},
  year = {2020},
}

@article{Fedotov-PRE-2010,
  title = {Non-Markovian random walks and nonlinear reactions},
  author = {Fedotov, Sergei},
  journal = {Phys. Rev. E},
  volume = {81},
  pages = {011117},
  year = {2010},
}

@article{Farago-2024,
  title = {Confined run-and-tumble particles with non-Markovian tumbling statistics},
  author = {Farago, Oded and Smith, N. R.},
  journal = {Phys. Rev. E},
  volume = {109},
  pages = {024110},
  year = {2024},
}

@article{PRE-FrydelPodgornik-2023,
  author  = {Frydel, Derek and Podgornik, Rudi},
  title   = {Mean-field theory of active electrolytes: Dynamic adsorption and overscreening},
  journal = {Phys. Rev. E},
  year    = {2023},
  volume  = {107},
  number  = {2},
  pages   = {024603},
  doi     = {10.1103/PhysRevE.107.024603}
}

@article{PRL-FilyMarchetti-2012,
  author  = {Fily, Yaouen and Marchetti, M. Cristina},
  title   = {Athermal Phase Separation of Self-Propelled Particles with No Alignment},
  journal = {Phys. Rev. Lett.},
  volume  = {108},
  number  = {23},
  pages   = {235702},
  year    = {2012},
  doi     = {10.1103/PhysRevLett.108.235702}
}

@article{RMP-Bechinger-2016,
  author  = {Bechinger, Clemens and Di Leonardo, Roberto and L{\"o}wen, Hartmut and Reichhardt, Charles and Volpe, Giorgio and Volpe, Giovanni},
  title   = {Active particles in complex and crowded environments},
  journal = {Rev. Mod. Phys.},
  volume  = {88},
  number  = {4},
  pages   = {045006},
  year    = {2016},
  doi     = {10.1103/RevModPhys.88.045006}
}

@article{SparreAndersen1953,
  author  = {Sparre Andersen, E.},
  title   = {On the fluctuations of sums of random variables},
  journal = {Mathematica Scandinavica},
  volume  = {1},
  year    = {1953},
  pages   = {263--285}
}

@article{SparreAndersen1954,
  author  = {Sparre Andersen, E.},
  title   = {On the fluctuations of sums of random variables II},
  journal = {Mathematica Scandinavica},
  volume  = {2},
  year    = {1954},
  pages   = {195--223}
}

@article{PRL-Klinger-2022,
  title = {Splitting Probabilities of Symmetric Jump Processes},
  author = {Klinger, J. and Voituriez, R. and B\'enichou, O.},
  journal = {Phys. Rev. Lett.},
  volume = {129},
  issue = {14},
  pages = {140603},
  numpages = {6},
  year = {2022},
  month = {Sep},
  publisher = {American Physical Society},
  doi = {10.1103/PhysRevLett.129.140603},
  url = {https://link.aps.org/doi/10.1103/PhysRevLett.129.140603}
}

@article{Malakar-2018,
doi = {10.1088/1742-5468/aab84f},
url = {https://dx.doi.org/10.1088/1742-5468/aab84f},
year = {2018},
month = {apr},
publisher = {IOP Publishing and SISSA},
volume = {2018},
number = {4},
pages = {043215},
author = {Kanaya Malakar and V Jemseena and Anupam Kundu and K Vijay Kumar and Sanjib Sabhapandit and Satya N Majumdar and S Redner and Abhishek Dhar},
title = {Steady state, relaxation and first-passage properties of a run-and-tumble particle in one-dimension},
journal = {Journal of Statistical Mechanics: Theory and Experiment}
}

@article{PRL-Cates-2008,
  title = {Statistical Mechanics of Interacting Run-and-Tumble Bacteria},
  author = {Tailleur, J. and Cates, M. E.},
  journal = {Phys. Rev. Lett.},
  volume = {100},
  issue = {21},
  pages = {218103},
  numpages = {4},
  year = {2008},
  month = {May},
  publisher = {American Physical Society},
  doi = {10.1103/PhysRevLett.100.218103},
  url = {https://link.aps.org/doi/10.1103/PhysRevLett.100.218103}
}

@article{PRE-Schnitzer-1993,
  title = {Theory of continuum random walks and application to chemotaxis},
  author = {Schnitzer, Mark J.},
  journal = {Phys. Rev. E},
  volume = {48},
  issue = {4},
  pages = {2553--2568},
  numpages = {0},
  year = {1993},
  month = {Oct},
  publisher = {American Physical Society},
  doi = {10.1103/PhysRevE.48.2553},
  url = {https://link.aps.org/doi/10.1103/PhysRevE.48.2553}
}

@article{POF-Frydel-2024,
    author = {Frydel, Derek},
    title = "{Run-and-tumble particles in slit geometry as a splitting probability problem}",
    journal = {Physics of Fluids},
    volume = {36},
    number = {11},
    pages = {111901},
    year = {2024},
    month = {11},
    doi = {10.1063/5.0233462},
    url = {https://doi.org/10.1063/5.0233462},
}

@article{JSTAT-Frydel-2021,
doi = {10.1088/1742-5468/ac1665},
url = {https://dx.doi.org/10.1088/1742-5468/ac1665},
year = {2021},
month = {aug},
publisher = {IOP Publishing and SISSA},
volume = {2021},
number = {8},
pages = {083220},
author = {Derek Frydel},
title = {Generalized run-and-tumble model in 1D geometry for an arbitrary distribution of drift velocities},
journal = {Journal of Statistical Mechanics: Theory and Experiment}
}

@article{pressure-2015b,
  title = {Pressure and Phase Equilibria in Interacting Active Brownian Spheres},
  author = {Solon, Alexandre P. and Stenhammar, Joakim and Wittkowski, Raphael and Kardar, Mehran and Kafri, Yariv and Cates, Michael E. and Tailleur, Julien},
  journal = {Phys. Rev. Lett.},
  volume = {114},
  issue = {19},
  pages = {198301},
  numpages = {6},
  year = {2015},
  month = {May},
  publisher = {American Physical Society},
  doi = {10.1103/PhysRevLett.114.198301},
  url = {https://link.aps.org/doi/10.1103/PhysRevLett.114.198301}
}

@book{Redner2001,
  title     = {A Guide to First-Passage Processes},
  author    = {Sidney Redner},
  year      = {2001},
  publisher = {Cambridge University Press},
  address   = {Cambridge},
  isbn      = {978-0521652244}
}

\end{document}